\documentclass[12pt]{article}
\usepackage{amsmath} 
\usepackage{graphicx}
\usepackage{bm}
\usepackage{times} 
\usepackage{braket} 
\usepackage{color,graphicx}
\usepackage{slashed} 
\usepackage{hyperref}

\newcommand{\qq}{q^2}

\newcommand {\E}[1]{\times 10^{#1}} 
\newcommand {\e}[1]{\mathrm{~#1}} 
\newcommand{\mc}[1]{\mathcal{#1}}

\newcommand{\mrm}[1]{\mathrm{#1}}

\renewcommand{\Re}[0]{\mrm{Re}}
\renewcommand{\Im}[0]{\mrm{Im}}

\def\pbnr{}
\def\speaker{Svjetlana Fajfer}
\def\title{Theory of rare charm decays}
\def\affiliation{J. Stefan Institute, Jamova 39, P. O. Box 3000, 1001 Ljubljana, Slovenia and\\
Faculty of Mathematics and physics, University of Ljubljana, Jadranska 19, 1000 Ljubljana, Slovenia}
\def\support{The workshop was supported by the University of Manchester, IPPP, STFC, and IOP}

\newcommand\pubnumber{\pbnr}
\newcommand\pubdate{\today}
\def\Title#1{\begin{center} {\Large #1 } \end{center}}
\def\Author#1{\begin{center}{ \sc #1} \end{center}}

\newcommand{\OnBehalf}[1]{\sbox0{#1}\ifdim\wd0=0pt
        {}
	\else
	{\\on behalf of #1}
	\fi}
\newcommand{\SupportedBy}[1]{\sbox0{#1}\ifdim\wd0=0pt
        {}
	\else
	{\footnote{#1}}
	\fi}
\def\Address#1{\begin{center}{ \it #1} \end{center}}

\newcommand\pubblock{\includegraphics[width=5cm]{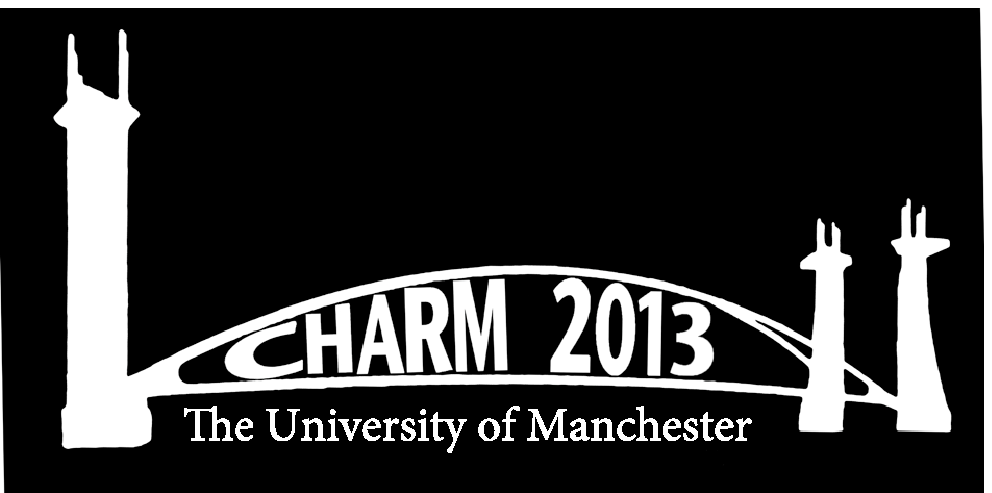}\hfill{\begin{tabular}{l} \pubnumber\\
         \pubdate  \end{tabular}}}
\newenvironment{Abstract}{\begin{quotation}  }{\end{quotation}}
\newenvironment{Presented}{\begin{quotation} \begin{center} 
             PRESENTED AT\end{center}\bigskip 
      \begin{center}\begin{large}}{\end{large}\end{center} \end{quotation}}
\def\Acknowledgements{\bigskip  \bigskip \begin{center} \begin{large}
             \bf ACKNOWLEDGEMENTS \end{large}\end{center}}
\def\venue{The 6$^{th}$ International Workshop on Charm Physics\\
(CHARM 2013)\\
Manchester, UK,  31 August -- 4 September, 2013}

\def\beq{\begin{equation}}
\def\eeq#1{\label{#1}\end{equation}}
\def\eeqn{\end{equation}}

\def\beqa{\begin{eqnarray}}
\def\eeqa#1{\label{#1}\end{eqnarray}}
\def\eeqan{\end{eqnarray}}

\let\bar=\overbar

\def\Dslash{\not{\hbox{\kern-4pt $D$}}}
\def\dslash{\not{\hbox{\kern-2pt $\del$}}}

\def\msb{{\bar{\ssstyle M \kern -1pt S}}}

\begin{document}
\begin{titlepage}
\pubblock

\vfill
\Title{\title}
\vfill
\Author{\speaker\SupportedBy{\support}}
\Address{\affiliation}
\vfill
\begin{Abstract}
Recent disagreement between experimental measurments of CP violating asymmetry in $D^0 \to K^+ K^-$ and $D^0 \to \pi^+ \pi^-$ and theoretical Standard model  expectation 
motivated many studies within the Standard model and beyond. Rare charm decays offer new probe of possible signals beyond Standard model.
CP conserving and CP violating contributions within the Standard model and beyond are reviewed for inclusive $c \to u \gamma$ and $c \to u l^+ l^-$ and exclusive $D \to V\gamma$, $D^0  \to P^+ P^-$, $D^+ \to \pi^+ l^+ l^-$,  $D \to l^+ l^-$ decays.

\end{Abstract}
\vfill
\begin{Presented}
\venue
\end{Presented}
\vfill
\end{titlepage}
\def\thefootnote{\fnsymbol{footnote}}
\setcounter{footnote}{0}
%

\section{Introduction}
For more than two decades up-quark sector was not considered as interesting testing ground for new physics searches. 
The  non-leptonic D mesons decay dynamics is polluted by the presence of many non-charm resonances in the vicinity of D meson masses. 
Flavour changing neutral current (FCNC) processes  are specially interesting  in searches of new physics. In the case of charm rare decays 
GIM mechanism plays special role. The interplay of CKM parameters and masses of down-like quarks leads to strong suppression in all  rare D decays.
The long distance contributions overshadow the short distance effects. The main issue is how to separate  information on short distance dynamics, either within SM or in its extensions. 
This is  a longstanding problem in rare charm decays.
Discrepancy between measured and expected CP violating asymmetry in charm decays \cite{Aaij:2011in,Aaltonen:2011se, HFAG} triggered many studies  of additional checks of the observed anomaly. 
Many theoretical studies were performed in order to explain the observed discrepancy \cite{Isidori:2011qw,Grossman:2006jg,Cheng:2012wr,Li:2012cfa,Franco:2012ck,Pirtskhalava:2011va,Bhattacharya:2012ah,Brod:2012ud,Feldmann:2012js}. Some of these approaches  have  explained  observed asymmetry by  the Standard model effects 
\cite{Brod:2012ud}, while in the rest of them, possible new physics effects were considered. In \cite{Isidori:2011qw} and \cite{Altmannshofer:2012ur} it was pointed out that most likely effective operators explaining CP asymmetry in charm decays are   color-magnetic dipole operators.  Most important result of these studies \cite{Isidori:2011qw,Grossman:2006jg,Cheng:2012wr,Li:2012cfa,Franco:2012ck,Pirtskhalava:2011va,Bhattacharya:2012ah,Feldmann:2012js} is that apparently one needs additional source of CP violation and in particular in the charm sector. 
The relevant question   is: Is there any possibility to observe CP violation in charm rare decays?
In Sec. 2  contributions to  $c \to u \gamma$ and $c \to u l^+ l^-$  decay modes are  reviewed. The exclusive radiative weak charm meson decays and possibility to search for CP violation are discussed in Sec. 3. Tests of CP violation in charm meson decays with the leptons in the final state  are discussed in Sec. 4.
Last section  contains  the summary.

\section{Inclusive decay modes: $c \to u \gamma$ and $c \to u l^+ l^-$ }
The effective low-energy Lagrangian describing $c \to u \gamma$ and $c \to u l^+ l^-$  transitions within SM is given by:
\begin{equation}
{\cal L}^{SD}_{eff} = \frac{G_F}{\sqrt 2} V_{cb}^* V_{ub} \sum_{i=7,9,10} C_i Q_i,
\label{e1}
\end{equation}
The operators are then:
\begin{eqnarray}
Q_7 &=&\frac{e}{8 \pi^2} m_c F_{\mu \nu} \bar u \sigma_{\mu \nu}(1+\gamma_5) c,\nonumber\\
Q_9 &=&\frac{e^2}{16 \pi^2} \bar u_L \gamma_{\mu } c_L \bar l \gamma^\mu l,\nonumber\\
Q_{10} &=&\frac{e^2}{16 \pi^2} \bar u_L \gamma_{\mu } c_L \bar l \gamma^\mu \gamma_5 l.\label{e2}
\end{eqnarray}
In (\ref{e1}) $C_i$ denote as usual Wilson coefficients  (they are determined at the scale $\mu = m_c$, $F_{\mu \nu}$ is the electromagnetic field strenght and $q_L = \frac{1}{2} ( 1-\gamma_5) q$. 
In the case of $c \to u \gamma$ decay only $C_7$ contributes, while in the case of $c \to u l^+ l^-$ all three Wilson coefficents are present. 
 At one-loop level contributions coming from penguin diagrams  is strongly GIM suppressed giving a branching ratio $\sim 10^{-18}$~\cite{Burdman:1995te,Fajfer:1998dv,Fajfer:1998rz,Fajfer:2001sa,Fajfer:2002gp}. 
 The QCD corrections enhance this rate to  $BR(c \to u \gamma )_{SM} =2.5 \times  10^{-8}$ ~\cite{Greub:1996wn,HoKim:1999bs}. Within Standard model the short distance contribution coming from $Q_{7,9}$  leads to the branching ratio \cite{Fajfer:2001sa,Burdman:2001tf,Paul:2011ar} 
 \begin{equation}
BR(D\to X_u e^+ e^-)^{SD}_{SM} \simeq 3.7 \times 10^{-9}.
\label{e3}
\end{equation}
However, this short distance contribution is overshadowed by long distance contributions \cite{Fajfer:2001sa,Burdman:2001tf} of the size:
\begin{equation}
BR(D\to X_u e^+ e^-)^{LD}_{SM} \sim {\cal O}(10^{-6}).
\label{e4}
\end{equation}
One of the most popular  extension of the SM  is MSSM. Following discussion in \cite{Fajfer:2007dy} the model with non-universal soft-breaking terms, knowing that gluino (according to LHC bounds)  cannot be lighter than $1.3$ ${\rm TeV}$, would give $Br(c \to u \gamma)_{gluino}\sim 5 \times  10^{-8}$. Rather high mass of gluino would also give  rise to SM $BR(c \to u l^+ l^-)$ by about factor 2. 
As noticed in \cite{Fajfer:2007dy} the other SM extensions might give larger increase of both inclusive branching ratios. However, regardless of the increase of short distance contribution the long distance effects screen their effect in the exclusive charm meson decay modes.\\

\section{Exclusive decay modes: $D \to V \gamma$ and $D \to P^+ P^- \gamma$}

The amplitude for the $D \to V \gamma$ decay  can be written as 
\begin{eqnarray}
{\cal A}[ D(p) \to V(p^\prime \epsilon^{\prime} \gamma(q, \epsilon)] &=& -i A_{CP} \epsilon_{\mu\nu\alpha \beta} q^\mu \epsilon^{*\nu} p^\alpha \epsilon^{*\prime, \beta}\nonumber\\
& +&A_{PV}[ (\epsilon^{*\prime, \beta}\cdot q)( \epsilon^{*\nu}\cdot q)
-(p\cdot q) (\epsilon^{*\nu}\ \epsilon^{*\nu})]\,. 
\label{e5}
\end{eqnarray}
In  the recent analysis \cite{Isidori:2012yx} the authors have reinvestigated  long distance dynamics. Using QCD sum rules result for the tensor form factors ($T^\rho \simeq T^\omega \simeq 0.7\pm 0.2$ from  they found
\begin{equation}
 (A_{PC,PV}^{\rho,\omega})^{SD} \simeq \frac{0.6(2) \times 10^{-9} }{m_D} \mid \frac{C_7(m_c)}{0.4\cdot 10^{-2}}\mid
\label{e6}
\end{equation}
where superscripts $\rho,\omega$ denote appropriate vector meson state $V$.  For the determination of short distance contribution one has to know matrix element of the $Q_7$ operator. In the calculations of it the tensor form-factors are present.  In ref.  \cite{Wu:2006rd} QCD sum rules were used to determine its structure. 
The long distance contribution was estimated by knowing that the relation $BR(D^0 \to K^{*0 }\gamma)/ BR(D^0 \to K^{*0 }\rho^0) = BR(D^0 \to \phi \gamma)/ BR(D^0 \to \phi \rho^0) $ is a consequence of vector meson dominance 
 \cite{Isidori:2012yx}
\begin{eqnarray}
|(A_{PC,PV}^V)^{LD} |= [ \frac{32 \pi}{2 m_D^3}  (1- \frac{m_V^2}{m_D^2}) ^{-3} \Gamma (D \to V \gamma) ]^{1/2},
\label{e7}
\end{eqnarray} 
what gives,  for $V=\phi$,   $| (A_{PC,PV}^\phi)^{LD}| =  \frac{5.9(4) \times 10^{-8}}{m_D} $. The main concern of this study was to investigate CP asymmetry in these decay modes.
The CP asymmetry is defined as 
\begin{eqnarray}
a_f\equiv \frac{\Gamma (D^0 \to f) - \Gamma (\bar D^0 \to f) }{\Gamma (D^0 \to f) +\Gamma (\bar D^0 \to f) }\,.
\label{e8}
\end{eqnarray} 
A lot of attention has been paid to the recent measurments of the CP violating asymmetry in charm decays. The LHCb collaboration updated recently their analysis  leading to   decreased value of the world average CP asymmetry \cite{HFAG,Aaij:2011in,Aaltonen:2011se}, 
\begin{equation}
\Delta a_{CP} = (-0.329 \pm 0.121)\%\,.
\label{e9}
\end{equation}
Assuming that only NP generates such CP asymmetry,  the authors of  \cite{Isidori:2011qw}  noticed that most likely candidate, among effective operators which can explained deviation, is the chromo-magnetic operator  $Q_8$.  This operator, under QCD renormalization group can mix with the  electric dipole operator $Q_7$  \cite{Isidori:2012yx}. It results in the  fact that both Wilson coefficients $C_7$ and $C_8$ are of comparable size at the charm scale. In particular their imaginary parts are then 
 \begin{equation}
{\rm Im} [C_7^{NP}(m_c) ]| \simeq {\rm Im} [C_8^{NP}(m_c) ]| \simeq 0.02 \times 10^{-2}.
\label{e10}
\end{equation}
The imaginary part of $C_7^{SM} $ is  two orders of magnitude smaller. The NP contribution is comparable in size with the real part of SM $|C_7^{SM-eff }(m_c) |= (0.5 \pm 0.1) \times 10^{-2}$. This means that  if the phase of long distance contribution can be neglected, and relative strong phase is maximal, the CP asymmetry can reach  
 \begin{equation}
|a_{V\gamma}| \sim 5 \%\, .
\label{e11}
\end{equation}
The  amplitude for the $D \to P^+ P^- \gamma$ decay can be decomposed into \cite{Fajfer:2002bq}
 \begin{eqnarray}
 {\cal A} (D (P)\to P_1(p_1)  P_2(p_2)\gamma (q,\epsilon)) &= &\frac{G_F}{\sqrt 2} V_{ci}^*V_{uj} \{ F_0 \lbrack \frac{p_1 \cdot \epsilon}{p_1 \cdot  q}- \frac{p_2 \cdot \epsilon}{p_2 \cdot  q}\rbrack \nonumber\\
 +F_1  \lbrack(p_1 \cdot \epsilon) (p_2 \cdot  q) - (p_2 \cdot \epsilon) (p_1 \cdot  q) \rbrack &+ &F_2 \epsilon^{\mu \nu \alpha \beta} \epsilon_\mu P_\nu p_{1\alpha} p_{2\beta} \}\,. 
\label{e12}
\end{eqnarray} 
The first part is inner bremsstrahlung amplitude, the $F_1$  part  denotes  the electric transition and $F_2$ is  the  magnetic transition amplitude.  The differential decay width is then
 \begin{eqnarray}
\frac{d \Gamma}{ds} = \frac{m_D^3}{32 \pi} (1- \frac{s}{m_D^2} ) \frac{\sqrt s \Gamma_0}{\pi} [ | F_1(s)|^2 + | F_2(s)|^2\, .
\label{e13}
\end{eqnarray} 
 The electric and magnetic dipole transitions were determined assuming vector meson exchange, knowing decay width of $V \to PP$ and using Breit-Wigner formula for the resonances present in the amplitude  \cite{Isidori:2012yx}.  
Following ref.  \cite{Isidori:2012yx} one can find  that CP asymmetries for the region bellow and above  $\phi$ resonance:
 \begin{eqnarray}
|a_{K^+ K^- \gamma}|^{max} &\simeq &1\%,  \enspace  2m_K \le \sqrt s \le 1.05 \, {\rm GeV}, \nonumber\\
|a_{K^+ K^- \gamma}|^{max} &\simeq &3\%,  \enspace  1.05\le \sqrt s \le 1.20 \, {\rm GeV} \,. 
\label{e14}
\end{eqnarray} 
In Table 1, the recent or existing estimate of the branching ratios for $D$ rare decays are presented including the reference. 

\begin{table}[t]
\begin{center}
\begin{tabular}{| l |c |c|}  
 \hline
Decay mode & Branching ratio &   Reference \\ \hline
 $D\to \rho(\omega)\gamma$  &   $0.6\times10^{-5}$ &   1210.6546; 1205.3164  \\ 
 $D\to K^+K^-\gamma$  &   $1.35\times10^{-5}\, (\phi )$ &    1205.3164  \\  
 $D\to X_u l^+ l^-$  &   ${\cal O} (10^{-6})$ &    1101.6053 \\
 $D^+ \to \pi^+ l^+ l^-$  &   $2 \times 10^{-6}$ &    1208.0795; 0706.1133\\
 $D^+_s  \to K^+ l^+ l^-$  &   $6 \times 10^{-6}$ &    0706.1133  \\
 $D\to \pi^+ K^-  l^+ l^-$  &   ${\cal O} (10^{-5})$ &    1209.4253 \\
  $D\to \pi^+ \pi^-  l^+ l^-$  &   ${\cal O} (10^{-6})$ &    1209.4253 \\
 $D\to K^+ K^-  l^+ l^-$  &   ${\cal O} (10^{-7})$ &    1209.4253 \\
$D\to \pi^- K^+  l^+ l^-$  &   ${\cal O} (10^{-8})$ &    1209.4253 \\
 $D\to \gamma \gamma$  &   $ (1-3)\times10^{-8}$ &   1008.3141 \\ 
  $D\to \mu^+ \mu^-$  &   $ (7-8)\times10^{-13}$ &   1008.3141; 0903.3650 \\ 
 \hline
\end{tabular}
\caption{Branching ratios for charm meson decays.  The second column  contains  the SM theoretical predictions in which long- distance contribution is dominant. The last column contains the most recent references. }
\label{tab:BR}
\end{center}

\end{table}

\section{ Rare  $D^+ \to \pi^+ \mu^+ \mu^-$, $ D\to  hh l^+ l^-$  decays and CP violation }

In this section  rare $D \to \pi  \ell^+ \ell^-$  decays are  reviewed with the goal to  determine  possibility  to study CP violation observables  \cite{Fajfer:2012nr}. 
 The new CP violating
  effects in rare decays $D \to P \ell^+ \ell^-$ might arise due to
  the interference of resonant part of the long distance contribution
  and the new physics affected short distance contribution. 
    The appropriate observables, 
  the differential direct CP asymmetry and partial decay width CP
  asymmetry are introduced  in a model independent way.  
  Among all decay modes the simplest one for the experimental searches are $D^+ \to \pi^+
  \ell^+ \ell^-$ and $D_s^+ \to K^+ \ell^+ \ell^-$.  The short distance dynamics for 
 $c \to u \ell^+ \ell^-$ decay on scale $\sim m_c$ is
discussed in details  by the effective
Hamiltonian given in ~\cite{Burdman:2001tf,Isidori:2011qw}. In the  decay width spectrum of
$c \to u \ell^+ \ell^-$ two light generations dominate short distance dynamics. 
Only when  third
generation is included  there is a possibility to obtain  non-vanishing imaginary
part: $\Im (\lambda_{b}/\lambda_d) = -\Im
(\lambda_{s}/\lambda_d)$. 
The CP violating parts of the
amplitude are suppressed by a very small factor $\lambda_b/\lambda_d \sim
10^{-3}$ with respect to the CP conserving ones and   therefore the CP violating effects should be very small. 
Due to the rather large  direct CP violation, measured in singly Cabibbo suppressed decays $D^0 \to \pi
\pi,KK$, one might expect similar increase in charm rare decays. 
If the CP violation arises due to new physics effects,  as it is mentioned already, it is due to 
the chromomagnetic operator $\mc{Q}_8$  contribution at some high
scale above $m_t$~\cite{Isidori:2011qw}. This as in the case of radiative weak decays comes from mixing of $\mc{O}_8$ into $\mc{O}_7$ under QCD renormalization.
 
 Close to the $\phi$ resonant peak the long distance amplitude for  $D^+ \to \pi^+ \mu^+ \mu^-$ decay is, to a
good approximation, determined  by non-factorizable contributions of
four-quark operators in $\mc{H}^s$. The width of $\phi$ resonance is
very narrow ($\Gamma_\phi/m_\phi \approx 4\E{-3}$) and well separated
from other vector resonances in the $\qq$ spectrum of $D \to P \ell^+
\ell^-$. Relying on vector meson dominance hypothesis the
$q^2$-dependence of the decay spectrum close to the resonant peak
follows the Breit-Wigner
shape~\cite{Burdman:2001tf,Fajfer:2005ke,Fajfer:2007dy}
\begin{equation}
\label{e15}
  \mc{A}_{\mathrm{LD}}^\phi \left[D  \to \pi  \phi \to  \pi \ell^- \ell^+\right]
   = \frac{i G_F }{\sqrt{2}} \lambda_s \frac{8\pi \alpha}{3}\, a_{\phi}  e^{i \delta_{\phi}}\,
   \frac{ m_{\phi} \Gamma_\phi   }{q^2-m_{\phi}^2 + i m_{\phi} \Gamma_{\phi}}\, \bar{u}(k_-)\, \slashed{p}\,v(k_+)\,.
\end{equation}
Finite width of the resonance generates a $q^2$-dependent strong phase
that varies across the peak. The strong phase 
on peak, $\delta_\phi$, and the normalization, $a_\phi$ are introduced in such a way, that both
are assumed to be independent of $q^2$. Parameter $a_\phi$ is real and can
be fixed from measured branching fractions of $D \to \pi \phi$ and
$\phi \to \ell^+ \ell^-$ decays~\cite{Fajfer:2007dy}.  We use  PDG values for 
$BR(D^+ \to \phi \pi^+) = (2.65 \pm 0.09 )\times 10^{-3}$, $BR(\phi \to \mu^+ \mu^-) = (0.287 \pm 0.019)\times 10^{-3}$, 
and  take into account the small width of $\phi$ by narrow width approximation as in \cite{Fajfer:2012nr}. 
With $\mc{A}_{LD}^\phi
= \bar{\mc{A}}_{LD}^\phi$ the differential direct CP violation becomes
\begin{eqnarray}
  \label{e16}
  a_{CP} (\sqrt{q^2}) &\equiv&
  \frac{|\mc{A}|^2-|\overline{\mc{A}}|^2}{|\mc{A}|^2+|\overline{\mc{A}}|^2}\nonumber\\
& =  &\frac{-3}{2\pi^2}   \frac{f_T(q^2)}{a_\phi} \frac{m_c}{m_D+m_\pi} 
\Im\left[ \frac{\lambda_b}{\lambda_s} C_7\right]
  \left[\cos \delta_\phi - \frac{q^2-m_\phi^2}{m_\phi \Gamma_\phi} \sin \delta_\phi
  \right] \,.
\end{eqnarray}
The imaginary part in the above expression can be approximated as
$\Im[\lambda_b C_7]/\Re \,\lambda_s$.  The  $\Im [\lambda_b C_7]$ was set  to be 
$0.8\E{-2}$ in order to illustrate largest possible CP
effect. Relative importance of the $\cos \delta_\phi$ and $\sin
\delta_\phi$ for representative choices of $\delta_\phi$ is shown on
the  plot in fig 1. of \cite{Fajfer:2012nr}. 
As presented   in \cite{Fajfer:2012nr} the CP asymmetry can be, depending on unknown $\delta_\phi$, even or odd with respect to the resonant peak position. 
 The asymmetry  can reach $a_{CP }\sim 1\%$ (see discussion in \cite{Fajfer:2012nr} .
 
In addition, a CP asymmetry of a partial width in the range
$m_1< m_{\ell\ell} < m_1$ can be introduced:
\begin{equation}
\label{e17a}
  A_\mrm{CP} (m_1,m_2) = 
  \frac{\Gamma(m_1< m_{\ell\ell} < m_2)-\bar \Gamma(m_1 < m_{\ell\ell} < m_2)}{\Gamma(m_1< m_{\ell\ell} < m_2)+\bar \Gamma(m_1 < m_{\ell\ell} < m_2)}\,,
\end{equation}
where $\Gamma$ and $\bar \Gamma$ denote partial decay widths of $D^+$
and $D^-$ decays, respectively, to $\pi^\pm \mu^+ \mu^-$.  $A_\mrm{CP}$
is related to the differential asymmetry $a_\mrm{CP}(\sqrt{q^2})$
as
\begin{equation}
  A_{CP}  (m_1,m_2) =
  \frac{\int_{m_1^2}^{m_2^2} dq^2 R(q^2) \,a_\mrm{CP} (\sqrt{q^2})}{\int_{q_\mrm{min}^2}^{q_{max}^2} dq^2 R(q^2)}\,,
  \label{e17b}
\end{equation}
where
\begin{equation}
  \label{eq:13}
  R(q^2) = \frac{1}{(q^2-m_\phi^2)^2+m_\phi^2 \Gamma_\phi^2}
  \int_{s_\mrm{min}(q^2)}^{s_\mrm{max}(q^2)} ds\, \sum_{s_+,s_-} \left|\bar u^{(s_-)}(k_-)\, , 
  \,\slashed{p} \,v^{(s_+)}(k_+)\right|^2\, ,
\end{equation}
involves the resonant shape and the
integral of the lepton trace over the Dalitz variable as in  \cite{Fajfer:2012nr}. 
The asymmetry on the same bin for the $\pi^+ \mu^+ \mu^-$ final state can be defined as 
\begin{equation}
\label{e17}
  C^{\phi}_\mrm{CP} \equiv A_{CP} (m_\phi - 20 \e{MeV}, m_\phi + 20 \e{MeV})\,.
\end{equation}
The asymmetry $C^\phi_\mrm{CP}$ is most sensitive to the $\cos
\delta_\phi$. term.  Sensitivity   is therefore optimized for
cases when $\delta_\phi \sim 0$ or $\delta_\phi \sim \pi$.  Its
sensitivity would decrease if we approached $\delta_\phi \sim \pm\pi/2$, since the
$a_{CP}(m_{\ell\ell})$ would be asymmetric in $(m_{\ell\ell}-m_\phi)$ in
this case. For that
 region of $\delta_\phi$ it was found that  the following observable has  good
sensitivity to direct CP violation
\begin{eqnarray}
  \label{e18}
    S^{\phi}_\mrm{CP} &\equiv & A_{CP}(m_\phi - 40 \e{MeV}, m_\phi -    20 \e{MeV})\nonumber\\ 
    & -  & A_{CP}(m_\phi + 20 \e{MeV}, m_\phi + 40 \e{MeV})
\end{eqnarray}
The bins where the partial width CP asymmetries $C^{\phi}_\mrm{CP}$
and $S^{\phi}_\mrm{CP}$ are defined are shown in fig.~2 in  \cite{Fajfer:2012nr} 
together with $a_{CP}(m_{\ell\ell})$. The largest asymmetry $A_{CP}$ approaches $5\%$ for  $\delta_\phi =\pm\pi/2$. 
The detailed analysis of the  semileptonic four body $ D\to  hh l^+ l^-$ decays was done in the work  of ref. \cite{Cappiello:2012vg}. The dominant long-distance 
contributions (bremsstrahlung and hadronic effects) are calculated and  total branching ratios and  the ( $ m^2_{ll},m^2_{hh}$) Dalitz plots are presented.
Branching ratios   turn out to be substantially larger than previously expected. Using vector meson dominance, it was found  for the Cabibbo-allowed, singly Cabibbo-suppressed, and doubly Cabibbo-suppressed modes
\begin{eqnarray}
&&BR(D^0 \to  K^- \pi^+ l^+ l^-)  \sim 10^{-5}\nonumber\\
&&BR(D^0 \to \pi^- \pi^+ l^+ l^-)  \sim 10^{-6}\nonumber\\
&&BR(D^0 \to K^- K^+ l^+ l^-)  \sim 10^{-7}\nonumber\\
&&BR(D^0 \to K^+ \pi^- l^+ l^-)  \sim 10^{-8}\label{e19}
\end{eqnarray}
The   new physics detection in these decay modes was also discussed. It was found that  two angular asymmetries, namely the T-odd diplane asymmetry and the forward-backward dilepton asymmetry offer direct tests of new physics due to tiny Standard model backgrounds.  If supersymmetric and $Z^\prime$-enhanced scenarios are assumed,  and if the size of Wilson coefficients   $C_9$  and $C_{10}$  is compatible with the observed CP asymmetry in nonleptonic charm decays  and flavor constraints, it was found in \cite{Cappiello:2012vg} that new physics effects in $ D^0 \to h_1 h_2 l^+l^-$ might  reach the $\%$ level.
In Table 2  predictions for size of  CP violating asymmetries in rare charm decays are presented. 

The two body rare decays  $D^0 \to \gamma \gamma$  and $D^0\to  l^+l^-$  were reconsidered in \cite{Paul:2010pq}. The result for the short and long distance contributions are $BR_{2-loops}(D^0 \to \gamma \gamma)= (3.6 -8.1) \times 10^{-12}$.  Short distance contributions in $D^0\to  l^+l^-$ decay lead to a very suppressed branching ratio in the SM. Therefore, it is natural to consider it is as an ideal testing ground for NP effects. Ref. \cite{Paul:2010pq} considered contributions coming from  $\gamma \gamma$ intermediate states due to long distance dynamics in $D^0 \to \mu^+ \mu^-$  arriving at the value $BR(D^0 \to \mu^+ \mu^-)  \sim  (2.7 -   8) \times 10^{-13}$.  According to  calculations of the same authors,  some NP  models  can enhance the branching ratio by  a factor of 2.
Recently LHCb improved bound on the branching ratio $BR(D^0 \to \mu^+ \mu^-) \le 6.2\times10^{-9}$ \cite{Aaij:2013cza} and it  offers an ideal possibility to test NP models. 

\begin{table}[t]
\begin{center}
\begin{tabular}{| l |c |c|}  
 \hline
Decay mode &   size &Reference \\ 
\hline
 $D\to \rho(\omega)\gamma$  &   $\le 5\%$  &   1210.6546\\ 
 $D\to K^+K^-\gamma$   &  $\le 1\% (\le 3\%) $ &    1205.3164  \\  
 $D\to X_u l^+ l^-$  & $\le1\%$ &    1212.4849 \\
 $D^+ \to \pi^+ \mu^+ \mu^-$  &  $\le 5\% (1\%)$ &    1208.0795 \\
 $D^+ \to h h \mu^+ \mu^-$  &  $\le 1\%$&    1208.0795 \\
 \hline
\end{tabular}
\caption{CP violating asymmetries for charm rare decays, size  and the original reference.The four last decay modes have  the CP asymmetry in the vicinity  $\phi$ resonance. }
\label{tab:CP}
\end{center}
\end{table}

\section{Summary}
The SM contribution to rare charm decays are rather well known. For all decay modes amplitudes
 are fully dominated by long distance dynamics. 
The possible presence of CP violation induced by new physics  in charm nonleptonic decays 
open new window for  new physics searches.  
The study of rare charm decays were revived and number of studies of CP violation in rare charm decays were done.
Very interesting signals of new physics might arise in $D\to \rho(\omega)\gamma$  and $D\to K^+K^-\gamma$, as well as in decays with the leptonic pair in the final state 
$D\to X_u l^+ l^-$, $D^+ \to \pi^+ \mu^+ \mu^-$, $D^+ \to h h \mu^+ \mu^-$.  The three body decays decays are particularly interesting, since one can focus on the CP asymmetry around the $\phi$ resonant peak in
spectrum of dilepton invariant mass. The interference term between the resonant and the short distance
amplitude  drives the direct CP asymmetry.  If there is no enhancement of CP violation  in $D^+
\to \pi^+ \ell^+ \ell^-$ then one cannot judge whether CP violation in
$D \to \pi \pi, K K$ is entirely due to SM dynamics or not. 
Hovewer, by not observing  any CP asymmetry in $D^+ \to
\pi^+ \ell^+ \ell^-$ around the $\phi$ peak would suggest that  SM
explanation of the observed CP violation in $D \to \pi \pi,\, K K$ is most likely.
The study of CP violation in all rare charm decays might differentiate between possible explanations of the observed CP asymmetry in charm decays and  constrain 
 new physics in charm sector.




\Acknowledgements
I am grateful to Nejc Ko\v snik for his invaluable comments.


\begin{thebibliography}{99}

\bibitem{Aaij:2011in}
  R.~Aaij {\it et al.}  [LHCb Collaboration],
  Phys.\ Rev.\ Lett.\  {\bf 108} (2012) 111602
  [arXiv:1112.0938 [hep-ex]].
\bibitem{Aaltonen:2011se}
  T.~Aaltonen {\it et al.}  [CDF Collaboration],
  Phys.\ Rev.\ D {\bf 85} (2012) 012009
  [arXiv:1111.5023 [hep-ex]].
 \bibitem{HFAG} http://www.slac.stanford.edu/xorg/hfag/

\bibitem{Isidori:2011qw}
  G.~Isidori, J.~F.~Kamenik, Z.~Ligeti and G.~Perez,
  Phys.\ Lett.\ B {\bf 711} (2012) 46
  [arXiv:1111.4987 [hep-ph]].
\bibitem{Grossman:2006jg}
  Y.~Grossman, A.~L.~Kagan and Y.~Nir,
  Phys.\ Rev.\ D {\bf 75} (2007) 036008
  [hep-ph/0609178].
\bibitem{Cheng:2012wr}
  H.~-Y.~Cheng and C.~-W.~Chiang,
  Phys.\ Rev.\ D {\bf 85} (2012) 034036
   [Erratum-ibid.\ D {\bf 85} (2012) 079903]
  [arXiv:1201.0785 [hep-ph]].
\bibitem{Li:2012cfa}
  H.~-n.~Li, C.~-D.~Lu and F.~-S.~Yu,
  Phys.\ Rev.\ D {\bf 86} (2012) 036012
  [arXiv:1203.3120 [hep-ph]].
\bibitem{Franco:2012ck}
  E.~Franco, S.~Mishima and L.~Silvestrini,
  JHEP {\bf 1205} (2012) 140
  [arXiv:1203.3131 [hep-ph]].
\bibitem{Pirtskhalava:2011va}
  D.~Pirtskhalava and P.~Uttayarat,
  Phys.\ Lett.\ B {\bf 712} (2012) 81
  [arXiv:1112.5451 [hep-ph]].
\bibitem{Bhattacharya:2012ah}
  B.~Bhattacharya, M.~Gronau and J.~L.~Rosner,
  Phys.\ Rev.\ D {\bf 85} (2012) 054014
  [arXiv:1201.2351 [hep-ph]].
\bibitem{Brod:2012ud}
  J.~Brod, Y.~Grossman, A.~L.~Kagan and J.~Zupan,
  JHEP {\bf 1210} (2012) 161
  [arXiv:1203.6659 [hep-ph]].
\bibitem{Feldmann:2012js}
  T.~Feldmann, S.~Nandi and A.~Soni,
  JHEP {\bf 1206} (2012) 007
  [arXiv:1202.3795 [hep-ph]].
\bibitem{Altmannshofer:2012ur}
  W.~Altmannshofer, R.~Primulando, C.~-T.~Yu and F.~Yu,
  JHEP {\bf 1204} (2012) 049
  [arXiv:1202.2866 [hep-ph]].

\bibitem{Burdman:1995te}
  G.~Burdman, E.~Golowich, J.~L.~Hewett and S.~Pakvasa,
  Phys.\ Rev.\ D {\bf 52} (1995) 6383
  [hep-ph/9502329].
\bibitem{Greub:1996wn}
  C.~Greub, T.~Hurth, M.~Misiak and D.~Wyler,
  Phys.\ Lett.\ B {\bf 382} (1996) 415
  [hep-ph/9603417].
\bibitem{HoKim:1999bs}
  Q.~Ho-Kim and X.~-Y.~Pham,
  Phys.\ Rev.\ D {\bf 61} (2000) 013008
  [hep-ph/9906235].
\bibitem{Fajfer:1998dv}
  S.~Fajfer, S.~Prelovsek and P.~Singer,
  Eur.\ Phys.\ J.\ C {\bf 6} (1999) 471
  [hep-ph/9801279].
\bibitem{Fajfer:1998rz}
  S.~Fajfer, S.~Prelovsek and P.~Singer,
  Phys.\ Rev.\ D {\bf 58} (1998) 094038
  [hep-ph/9805461].
\bibitem{Fajfer:2001sa}
  S.~Fajfer, S.~Prelovsek and P.~Singer,
  Phys.\ Rev.\ D {\bf 64} (2001) 114009
  [hep-ph/0106333].
\bibitem{Fajfer:2002gp}
  S.~Fajfer, P.~Singer and J.~Zupan,
  Eur.\ Phys.\ J.\ C {\bf 27} (2003) 201
  [hep-ph/0209250].
\bibitem{Burdman:2001tf}
  G.~Burdman, E.~Golowich, J.~L.~Hewett and S.~Pakvasa,
  Phys.\ Rev.\ D {\bf 66} (2002) 014009
  [hep-ph/0112235].
\bibitem{Paul:2011ar}
  A.~Paul, I.~I.~Bigi and S.~Recksiegel,
  Phys.\ Rev.\ D {\bf 83} (2011) 114006
  [arXiv:1101.6053 [hep-ph]].
\bibitem{Fajfer:2005ke}
  S.~Fajfer and S.~Prelovsek,
  Phys.\ Rev.\ D {\bf 73} (2006) 054026
  [hep-ph/0511048].
\bibitem{Fajfer:2007dy}
  S.~Fajfer, N.~Kosnik and S.~Prelovsek,
  Phys.\ Rev.\ D {\bf 76} (2007) 074010
  [arXiv:0706.1133 [hep-ph]].
  \bibitem{Isidori:2012yx}
  G.~Isidori and J.~F.~Kamenik,
  Phys.\ Rev.\ Lett.\  {\bf 109} (2012) 171801
  [arXiv:1205.3164 [hep-ph]].
\bibitem{Wu:2006rd}
  Y.~-L.~Wu, M.~Zhong and Y.~-B.~Zuo,
  Int.\ J.\ Mod.\ Phys.\ A {\bf 21} (2006) 6125
  [hep-ph/0604007].

 \bibitem{HFAG-b}
arXiv:1010.1589 [hep-ex].
\bibitem{Fajfer:2002bq}
  S.~Fajfer, A.~Prapotnik and P.~Singer,
  Phys.\ Rev.\ D {\bf 66} (2002) 074002
  [hep-ph/0204306].
\bibitem{Fajfer:2012nr}
  S.~Fajfer and N.~Kosnik,
  Phys.\ Rev.\ D {\bf 87} (2013) 054026
  [arXiv:1208.0759 [hep-ph]].
\bibitem{Cappiello:2012vg}
  L.~Cappiello, O.~Cata and G.~D'Ambrosio,
  JHEP {\bf 1304} (2013) 135
  [arXiv:1209.4235 [hep-ph]].
\bibitem{Paul:2010pq}
  A.~Paul, I.~I.~Bigi and S.~Recksiegel,
  Phys.\ Rev.\ D {\bf 82} (2010) 094006
   [Erratum-ibid.\ D {\bf 83} (2011) 019901]
  [arXiv:1008.3141 [hep-ph]].
\bibitem{Aaij:2013cza}
  RAaij {\it et al.}  [LHCb Collaboration],
  Phys.\ Lett.\ B {\bf 725} (2013) 15
  [arXiv:1305.5059 [hep-ex]].

\end{thebibliography}
\end{document}